\documentclass[useAMS,usenatbib]{mn2e}

\title[Physics of radio emission]{Physics of radio emission in the long-period pulsars}
\author[P. B. Jones]{P. B. Jones\thanks{E-mail:
peter.jones@physics.ox.ac.uk}\\
University of Oxford, Department of Physics, Denys Wilkinson building,
Keble Road, Oxford OX1 3RH, England\\}
\begin{document}

\date{ }

\pagerange{\pageref{}--\pageref{}} \pubyear{}

\maketitle

\label{firstpage}

\begin{abstract}

Recent multi-frequency measurements of pulse widths $W_{50}$ for the long-period pulsar J0250+5854 by Agar et al provide a unique insight to the emission process owing to its small polar-cap radius. The frequency dependence of $W_{50}$ can be simply understood as a consequence of the emitting plasma remaining under acceleration during the interval of radio emission.  This is possible in a plasma of ions and protons but not in one of high-multiplicity electron-positron pairs. Extension of the model to the pulse profiles of the general pulsar population is considered briefly.

\end{abstract}

\begin{keywords} 
 pulsars: general: individual J0250-5854 - plasma - instabilities

\end{keywords}

\section{Introduction}

A recent paper by Agar et al (2021) describes the frequency dependence of the full-width at half-maximum (FWHM) emission profile of the 23.5 s pulsar J0250+5854 first observed by Tan et al (2018). This is remarkable in that the FWHM varies little between 149 and 350 MHz but then increases by more than a factor two at 1250 MHz: a change that is the opposite of that expected if the natural divergence of open flux lines were the dominant source of profile width. It offers a new insight into the angular distribution of radio emission with respect to the local magnetic flux density ${\bf B}$ at source, a factor usually neglected in studies of the general pulsar population.  J0250+5854 is expected to have the smallest polar cap of any known radio pulsar: if circular, the radius would be $u_{0}(1) = 3.4 \times 10^{3}$ cm.  The tangent to the flux lines at the open magnetosphere boundary would be $3u_{0}(1)\eta^{1/2}/2R = 0.24\eta^{1/2}$ degrees with respect to the magnetic axis in an undisturbed dipole field, where $\eta$ is the radius in units of $R = 1.2 \times 10^{6}$ cm, the neutron-star radius.

 Presumably, the radius $u_{0}(\eta)$ of the long narrow tube forming the open sector of the magnetosphere enclosing flux lines that intersect the light cylinder must be large enough to permit the growth of a longitudinal collective plasma mode, a width at least of the order of a mode wavelength in the rest frame of the emitting particles. The Lorentz factor of this frame must be no more than of the order of $10^{1-2}$ for an adequate mode growth rate.  Consequently, the distribution of the source angle $\chi$ can be expected to make a discernible contribution to the J0250+5854 FWHM.  This is the subject of the present letter.  Surprisingly, the source angular distribution can also be significant for the general pulsar population. Values of basic parameters used here have been taken from Manchester et al (2005).

\section{The emitting system}

 The basis of the ion-proton model (Jones 2012a,b, 2020) in pulsars with positive polar-cap corotational charge density (Goldreich \& Julian 1969) is an outgoing plasma of two components: protons and ions. These have different charge-to-mass ratios and hence different $\delta$-function velocities at any altitude above the polar-cap surface. They are accelerated from the the compact neutron-star atmosphere and are unstable against a Langmuir longitudinal or quasi-longitudinal mode whose growth exponent is approximately,
\begin{eqnarray}
\Lambda(\eta) = 7\times  10^{7}\frac{R}{c}\left(\frac{B_{12}}{P}\right)^{1/2}
  \int^{\eta}_{1} d\eta ( \eta \gamma)^{-3/2}
\end{eqnarray}
where $B_{12}$ is the surface field in units of $10^{12}$ G and $\gamma(\eta)$ is here the proton Lorentz factor.  It is assumed that $\Lambda_{c} = 30$ enables growth to non-linearity and our assumption is that, as a consequence of the non-uniformity of $\gamma$ across the width of the open sector, the form of the ensuing turbulence allows coupling with the radiation field which would not otherwise be possible for a longitudinal mode.  The unscreened potential accelerating the beams is derived from the Lense-Thirring effect (Muslimov \& Tsygan 1992; Harding \& Muslimov 2001) and is $\Phi(\eta) \approx 47(1 - \eta^{-3})$ GeV for J0250+5854.  Evaluation of equation (1) gives $\Lambda = \Lambda_{c}$ at $\eta \approx 1.1$.  Coupling with the radiation field moves the system towards its lowest energy state in which the two particle species have a common velocity.  But because both particles are positively charged, the potential $\Phi$ continues to accelerate the turbulent system, widening the velocity difference and increasing Lorentz factors.  (This is an essential difference from an electron-positron plasma of even moderate multiplicity which must exist in a zero ${\bf E}_{\parallel}$ region and cannot be accelerated.)

 Lorentz factors and $\Phi$ are modified by the screening effect of the reverse flux of photoelectrons from the accelerated ions. The ions are initially in local thermal equilibrium in the neutron star atmosphere but the Lorentz transformation of thermal photons to the accelerated ion frame and the large photoelectric cross-sections produce further ionization.  The electrons are accelerated back to the surface and are a source of protons which are produced in electromagnetic showers, by decay of the nuclear giant-dipole state, at the neutron-star surface at a rate of approximately one per $5$ GeV electron  energy.  The degree of screening depends on the atomic number of surface nuclei and, principally, the whole-surface temperature of the neutron.   The polar-cap temperature may well be higher than that of the whole surface, but the polar cap has little solid angle in the frame of the accelerated ion and Lorentz transformations from it to that frame are less likely to provide the photon energy boost necessary for ionization.  Reverse electrons with high Lorentz factors have negligible effect on the Langmuir mode dispersion relation. The role of these electrons in screening is entirely analogous with screening by pair creation.  It leads to a self-regulating acceleration potential in the general pulsar population several orders of magnitude smaller than the unscreened potential. The atomic number of surface nuclei and the neutron-star whole-surface temperature principally determine details of the screening but are are essentially unknown, as is the precise nature of the turbulence.

We are obliged to assume that in the turbulent plasma rest-frame, the radiation is isotropic with photon number spectrum
$\propto (\nu_{c0}/\nu_{c})^{\alpha}$ for $\nu_{c} > \nu_{c0}$ in which the turbulence evolves towards higher wave-numbers, as it does more generally, for example, in isotropic homogeneous fluids.  Here $\nu_{c}$ is the rest-frame frequency and $\nu_{c0}$ is a constant approximately equal to the local mode frequency.  But there is a further complication: the outward-moving plasma expands laterally and also tends to evolve adiabatically to smaller wave-number components.  It has to be admitted that the details of these stages of radiation are quite unclear.  But the radiation intensity as a function of $\chi$ in the observer frame, per unit frequency interval, solid angle and time is $\propto (1 + \gamma^{2}\chi^{2})^{-\alpha -2}$ for small $\chi$ (Jones 2017).  Setting $\alpha = 3.5$, with FWHM = $2\chi$, we find from the FWHM given by Agar et al, $\chi = 0.021$ and $\gamma = 18$ at 1250 MHz, and $\chi =0.010$ and $\gamma = 37$ at 149 MHz. These assume the FWHM derives entirely from the source angular distribution, neglecting the contribution from flux-line curvature which is here smaller than $0.008 \eta^{1/2}$ radians.  We might suppose that the formation of turbulence occurs within outward motion through an observer-frame length interval of the order of 20 open-sector radii, $(\sim 7 \times 10^{4}\eta^{3/2}$ cm).  Using the unscreened potential $\Phi$ we find $\gamma = 18$ at $\eta = 1.16$ and $\gamma = 37$ at $\eta = 1.57$, the difference being a distance of approximately $5 \times 10^{5}$ cm.  The complete emission process occurs within an interval of about $10^{6}$ cm.

We emphasize that specific numerical values given here should not be taken too seriously owing to our lack of knowledge of the parameters on which they depend.  Whole-surface temperatures $T_{s}$ are not well known for older pulsars, excluding millisecond pulsars for which there are other considerations (see the survey given by Kantor \& Gusakov, 2021).  Hubble Space Telescope observations have produced an interval $T_{s} = 1.3 - 2.5 \times 10^{5}$ K for B0950+08 (age 17.5 Myr; Pavlov et al 2017), $T_{s} = 2.7 \times 10^{4}$ K for J0108-1431 (age 166 Myr; Abramkin et al 2021), and $T_{s} < 4.2 \times 10^{4}$ K for J2144-3933 (age 270 Myr; Guillot et al 2019).  It is not possible to make a firm temperature estimate for J0250+5854 (age 13.7 Myr) on the basis of these few results.

 Parenthetically, the unstable mode leading to radio emission would exist in J0250+5854 at $T_{s}$ too low for significant proton production by reverse photoelectrons if baryonic particles of different charge-to-mass ratios are naturally present and ionized at the polar-cap surface. Possible sources could be a residue of the neutron-star formation or a reverse flux of baryonic particles from the vicinity of the Y-point to the neutron-star surface.  In J0250+5854 the value of $\Phi$ is so small that $\Lambda = \Lambda_{c}$ can be reached even in the absence of electron screening. Examination of equation (1) shows that this would also be true for J2144-3933 owing to its similarly small acceleration potential $\Phi(\infty) = 28$ GeV and in view of the Guillot et al measurement it is clear that for this pulsar, the low-temperature mode of functioning must be operative.

The chemical potential of a classical magnetized electron gas of density $N_{e}$ is approximately $8.6T_{5}(\ln N_{e} - 60 - \ln B_{12})$ whilst the ground-state binding energy of neutral hydrogen is 200 eV at $B_{12} = 2$ increasing to 390 eV at $B_{12} = 26$ (see Harding \& Lai 2006). This indicates that atomic hydrogen would exist at the temperature found by Guillot et al, certainly outside the polar cap or any region close to it into which there is reverse flow of particles from the Y-point. We refer to the review of Contopoulos (2016) regarding this aspect of magnetospheric structure. Protons escaping from any hot region and transforming to atomic hydrogen diffuse easily in comparison with protons or ions.  On the polar cap the temperature $T_{pc}$ exceeds $T_{s}$ if there is even a very small input of energy from reverse particle flow: an input of about 100 MeV per Goldreich-Julian flux particle would raise $T_{pc}$ to $2\times 10^{5}$ K, enabling ionization of atomic hydrogen and subsequent acceleration.

\section{The general pulsar population}

Application of the ion-proton model to the general population has some differences from the long-period pulsars.  The average value of FWHM  observed at period $P =1$ s is $W_{50} = 6.2$ degrees (Pilia et al 2016) whereas derived solely from the flux-line divergence it would be only $2.4\eta^{1/2}$ degrees assuming the whole polar cap is active.  However, the whole-surface temperature is likely to be $T_{s} = 2 - 3\times 10^{5}$ K so that screening of the Lense-Thirring acceleration field is present at considerably lower ion Lorentz factors. For a typical spectral index $\alpha = 2$ we find, working as in the previous Section, $\gamma\chi = 0.44$ so that, as an example, for $\gamma =10$ 
the source width is $2\chi = 5.0$ degrees and actually exceeds the contribution from flux-line divergence. The complete FWHM at $\eta = 2$ would be $6.0$ degrees. Of course this figure is not a prediction but serves to demonstrate the consistency of the ion-proton model with observation.  Neglect of the source contribution would require the assumption of a much higher emission altitude, here $\eta = 7$, for consistency with observation.

\section{The pair creation problem}

It is not possible to avoid this problem in regard to the long-period pulsars, particularly J2144-3933, if negative polar-cap corotational charge density is assumed (see Young, Manchester \& Johnston 1999). The Guillot et al limit on $T_{s}$ excludes pair creation by the inverse Compton or Breit-Wheeler processes. Single-photon conversion of curvature radiation remains but with an ad hoc radius of curvature $\rho$. The sole condition that can be applied is that both photon creation and conversion should occur on open flux lines.  The photon momentum threshold $k$ for conversion is then given approximately by, $k\theta B_{12} > 6 mc$ in which $\theta^{2} = 2u_{0}(1)/\rho$.  Then the production of an initial pair is possible if,
\begin{eqnarray}
\rho < \left(\frac{3(2u_{0}(1))^{1/2}B_{12} \hbar\gamma^{3}}{12mc^{2}}\right)^{2/3}
\end{eqnarray}
where $\gamma$ is here the primary electron Lorentz factor, with $B$ and $\rho$ independent of $\eta$.  A posteriori evaluation of $\rho\theta$ confirms that our assumption of $\eta$-independence is satisfactory, also that the effect of neutron-star rotation is negligible. The electron value of $\gamma$ is derived here from the maximum possible acceleration potential which assumes the existence of a transitory vacuum interval in the open magnetosphere as in the model of Timokhin \& Arons (2013) also adopted by Philippov, Timokhin \$ Spitkovsky (2020).  A possible test for this  model in the millisecond pulsars has been described recently by Jones (2021).  The potential is approximately $\Phi(\infty) = 7 \times 10^{3}B_{12}/P^{2}$ Gev, and the upper limit for flux-line radius of curvature for J2144-3933 is $\rho < 3.0 \times 10^{5}$ cm.  This is merely the condition for creation of the first generation pair in a cascade. For J0250+5854 the limit is $3.3 \times 10^{6}$ cm. The canonical view is that these conditions should be ignored and that, in the general pulsar population, a pair plasma exists. We consider that the extension of this to the long-period pulsars is not reasonable, particularly for J2144-3933.  Also, as noted above, there is no way of accelerating an electron-positron plasma for consistency with the Agar et al observations.

 Neutron stars with both positive and negative polar-cap corotational charge density must presumably exist.  The ion-proton model considers only the positive case. Negative-case pulsars with values of $\Phi(\infty)$ and of
$\rho$ inadequate for pair creation, including millisecond and long-period pulsars, are not expected to be radio-loud because the primary electrons whilst unable to produce pairs would have Lorentz factors too high to participate in longitudinal collective motion.

\section{Conclusions}

The observed multi-frequency profiles of J0250+5854 are consistent with a single plasma source moving outwards on a bundle of flux lines within the open sector of the magnetosphere, the widths $W_{50}$ depending in the main part on the source angular width $\chi$ rather than the natural divergence of the flux lines.  To a lesser extent this would be true for J2144-3933 and the 12.1 s pulsar J2251-3711 (Morello et al, 2020), therefore multi-frequency profiles of these pulsars would be of interest.  Pulsars in the general population and millisecond pulsars with positive polar-cap corotational charge density have larger polar caps and can support several such plasma sources.  Recent multi-frequency observations of the general population reveal both decreasing and increasing values of $W_{10}$ and $W_{50}$ as functions of frequency in  substantial minorities of cases.  We refer to the most recent large-number study by Posselt et al (2021) who find the frequency-dependences are monotonic and also give a careful survey of this field. Surprisingly, as a result of much lower ion Lorentz factors in the general population, the source angular distribution can be at least of the same order as the flux-line  divergence.

In the ion-proton model, general-population pulsars are complex systems: there can be several sources active at any instant within a polar-cap having individual values of $\gamma$ and possibly of the ion atomic number and the ion-proton number ratio (see Jones 2020).  The screening depends on the temperature $T_{s}$ of that part of the whole neutron-star surface that is visible to ions accelerated from the polar cap.  The model is not capable of offering predictions for any particular pulsar: there are too many unknowns in what is a diverse population.  But it does offer some physical understanding and diversity of functioning and is apt for the observed general population.

\section*{Data availability}

The data underlying this work will be shared on reasonable request to the corresponding author.

\section*{Acknowledgments}

The author thanks the anonymous referee for comments which have significantly improved the presentation of this work.

\bsp

\label{lastpage}

\end{document}